\documentclass[3p,times,procedia]{elsarticle}
\usepackage{nupha_ecrc}
\usepackage{amsmath, amssymb}
\usepackage{mathrsfs}

\volume{00}

\firstpage{1}

\journalname{Nuclear Physics A}

\runauth{S. Floerchinger}

\jid{nupha}

\jnltitlelogo{Nuclear Physics A}

\usepackage{amssymb}

\usepackage[figuresright]{rotating}

\begin{document}

\begin{frontmatter}



\dochead{}

\title{Heavy ion collisions and cosmology}


\author{Stefan Floerchinger}

\address{Institut f\"{u}r Theoretische Physik, Philosophenweg 16, 69120 Heidelberg, Germany}

\begin{abstract}
There are interesting parallels between the physics of heavy ion collisions and cosmology. Both systems are out-of-equilibrium and relativistic fluid dynamics plays an important role for their theoretical description. From a comparison one can draw interesting conclusions for both sides. For heavy ion physics it could be rewarding to attempt a theoretical description of fluid perturbations similar to cosmological perturbation theory. In the context of late time cosmology, it could be interesting to study dissipative properties such as shear and bulk viscosity and corresponding relaxation times in more detail. Knowledge and experience from heavy ion physics could help to constrain the microscopic properties of dark matter from observational knowledge of the cosmological fluid properties.
\end{abstract}


\end{frontmatter}


\section{The analogy}
\label{}

The analogy between ultra-relativistic heavy ion collisions and cosmology has often been emphasized. What distinguishes the physics there from typical situations in condensed matter physics is the aspect of history or dynamical evolution. Experimentally, one cannot probe a static quark gluon plasma with external probes as it is possible for solid state systems. All information must be reconstructed from the final state, which contains essentially the decay products of the quark-gluon plasma. In cosmology, this is very similar. The cosmic microwave background allows to see back in time until photon decoupling but information about earlier times is not directly available. Similarly, one can observe the large scale structure today and, by looking at more distant galaxies, in the recent past, but large parts of the history of structure formation are not accessible.

On first sight there are big differences: cosmological and nuclear scales differ by many orders of magnitude, late time cosmology is dominated by gravity, QED and the still rather poorly understood physics of dark matter and dark energy, while heavy ion collisions are governed by QCD. Another difference is that only a single event can be studied in cosmology, in contrast to very many heavy ion collisions. 
However, from a more abstract viewpoint, there are many interesting parallels. In the theoretical description, a particularly nice one is that many aspects of the dynamics can be described in both cases by relativistic fluid dynamics. 

While the fluid produced in high energy nuclear collisions consists initially of a quark-gluon plasma which then expands and cools down and eventually undergoes a transition to hadronic degrees of freedom, the cosmological fluid has many more different stages. Very shortly after the big bang, there is also one epoch where it was dominated by quarks and gluons but this will actually not be in the focus of this talk. The first reason is that, unfortunately, only little information is transmitted to later times from this era, the other is that the biggest theoretical problems in current cosmology research actually arise from the latter epoch of structure formation. As I will argue, there are interesting parallels between heavy ion collision physics and late time cosmology although on a slightly more abstract level.

Before going deeper into the comparison, it is instructive to compare briefly the symmetries in both cases. They are actually of statistical nature: Concrete realizations break them but the statistical properties are symmetric. In cosmology the statistics arises from comparing different regions of space while in heavy ion collisions the statistical properties are for ensembles of events. Now, for the universe, the cosmological principle implies a three-dimensional translation and rotation symmetry whereas central heavy ion collisions have only a one-dimensional rotation symmetry and an approximate one-dimensional Bjorken boost invariance in the central rapidity region. Only the very central region in the transverse plane has also an approximate two-dimensional translation invariance.

From these considerations follow practical consequences: For the treatment of inhomogeneities or perturbations around an homogeneous and isotropic universe, it is very useful to work with three-dimensional Fourier transforms. Perturbations can be classified according to their transformation behavior with respect to rotations (into scalar, vector and tensor excitations) and translations (by a three dimensional wave vector). For heavy ion collisions, a Bessel-Fourier transform has similar advantages \cite{Floerchinger:2013rya, Floerchinger:2014fta} although not as far going as for cosmology because of less degree of symmetry in total.

\section{The problem of initial conditions}
\label{}

A problem shared by both cosmology and heavy ion physics is the one of initial conditions. Due to obvious reasons it is not possible to directly access the initial conditions for a fluid dynamic description in cosmology, say very shortly after the big bang, or for heavy ion collisions shortly after the collision.\footnote{One may argue that the situation is better for heavy ions because the wavefunction of nuclei that determines the initial state can be probed by other experiments such as proton-ion or electron-ion collisions.} Despite this principle problem, cosmology has evolved over the last decades into a precision science - a success that heavy ion physics would certainly like to follow. It is instructive to compare the treatment of initial conditions in both cases.

The main source of information in cosmology are perturbations. They can be classified according to their transformation properties with respect to rotations and translations into scalar, vector and tensor modes with different wave vectors. Vector modes and some scalar modes are decaying and need not be specified further. Tensor modes correspond to gravitational waves and can be neglected for many considerations. The most prominent role is played by growing scalar modes. For the relevant range of wavelengths it turns out that their statistical properties are close to Gaussian with an almost scale invariant initial spectrum, 
$$ \langle \delta({\bf k}) \, \delta({\bf k^\prime}) \rangle = P(k) \; \delta^{(3)}({\bf k} + {\bf k^\prime}), $$
with $P(k) = c \, k^{n_s-1}$. In this way, initial conditions for very many modes are actually specified by two numbers: the spectral index $n_s$ and the overall amplitude $c$. These are finally determined from observations. It is this simplicity or universality of initial conditions that allows to learn so much from the observed temperature spectrum of the cosmic microwave background and the large scale structure.

In contrast, in heavy ion collisions, the current state of the art are explicit realizations of initial conditions in terms of various Monte-Carlo models. These have typically many parameters and the resulting initial conditions, say for the energy density, usually have features at many different length scales. The question arises, whether the success of cosmology in characterizing initial conditions could be repeated here. This would imply to characterize statistical properties instead of explicit realizations and to focusing on the relevant modes in the relevant range of wavelengths. First steps in this direction have been made \cite{Floerchinger:2014fta, Teaney:2010vd, ColemanSmith:2012ka, Yan:2013laa, Bzdak:2013raa} but more progress seems possible.

\section{Theory of inhomogeneities and anisotropies}
\label{}

In cosmology, a very useful and powerful formalism exist for treating inhomogeneities: cosmological perturbation theory, see for example refs.\ \cite{Mukhanov:1990me, Weinberg:2008zzc}. This formalism, which has been developed over the years by many people, solves the evolution equations of a fluid coupled to gravity for perturbations around an homogeneous and isotropic background order-by-order. The linear approximation leads to a very detailed understanding how different modes evolve in different epochs, and of many interesting phenomena like horizon crossing, acoustic oscillations, Silk damping etc.
Besides the linear formalism, there is also a diagrammatic formalism to treat non-linear mode-mode couplings. In principle one could use a very similar theoretical approach to describe the evolution of perturbations in the fluid dynamic description of heavy ion collisions \cite{Floerchinger:2013rya, Floerchinger:2014cra}. The somewhat smaller degree of global symmetries increases somewhat the technical effort. On the other side, many complications do not arise because one does not have to deal with the gauge freedom of gravity. Also, more fields are relevant in cosmology and there are more different epochs that need to be treated separately.

One might think that a perturbative approach is much more useful for cosmology because the relative amplitude of fluctuations in the energy density with respect to the cosmological background is initially very small, $\Delta \epsilon / \bar \epsilon \ll 1$. In heavy ion collisions, the precise magnitude of initial perturbations is not well known but one may expect $\Delta \epsilon / \bar \epsilon \approx 1$.
During the cosmological history, scalar modes with wavelength inside the horizon start growing due to the attractive gravitational interaction and the associated gravitational collapse. A perturbative treatment ceases to be valid at late time for modes with small wavelength which have large relative amplitude, $\Delta \epsilon / \bar \epsilon > 1$. For the fluid dynamics of heavy ions, in contrast, no such late time instability is expected and the relative amplitude of perturbations decreases due to viscous effects. Numerical investigations for realistic viscosities indicate that non-linear effects remain within reach of a perturbative treatment during the time- and length scales of a heavy ion collision \cite{Floerchinger:2013tya, Brouzakis:2014gka}. Developing a perturbative description of anisotropies in heavy ion collisions in detail seems to be a well motivated goal for theory in the coming years. This would be helpful to gain a more detailed understanding but also to make the relation between little bangs and the big bang more explicit.

\section{Viscosities in cosmology}
\label{}

While both, cosmology and heavy ion collisions can be described to a large extend by relativistic fluid dynamics, there is one interesting difference in the theoretical setup: for heavy ion collisions, the viscosities are relatively small (with respect to the entropy density) but non-vanishing. In contrast, in cosmology one usually assumes an ideal fluid. At very early times, when the inhomogeneities are small, this makes presumably no or only a very minor difference. However, at late times the density contrast grows large and so do fluid velocity gradients. The dynamics could then be sensitive to dissipative effects.

An interesting question in this context is how non-vanishing shear and bulk viscosity would modify the formation of the large scale structure of the universe. One can easily see that the most important scalar modes are sensitive to the linear combination $\zeta+4 \eta /3$. Qualitatively, viscosities can slow down gravitational collapse and structure formation but in contrast to the effect of a non-vanishing pressure, structure is not actively ``washed out'' by viscosity. A more detailed discussion can be found in ref. \cite{Blas:2015tla}.

\subsection{Effective fluid properties from coarse graining}

Interestingly, for cosmology, two different types of viscosity can play a role. First, the standard viscosity is due to momentum transport by particles or radiation, i.\ e.\ the microscopic constituents of the fluid. It's value can be calculated close to equilibrium from linear response theory (Green-Kubo formula). In addition to this, there can be effects from the fact that the cosmological fluid is inhomogeneous and described in a statistical way. This adds another level of statistical fluctuations to the description, which goes beyond the fluctuating particle positions and momenta for given fluid velocity and density. More specific, in a coarse-grained description, where one averages over some of the statistical initial state fluctuations, the expectation values can follow evolution equations which resemble the form of the original fluid dynamics but with additional terms. These additional terms arising from the coarse graining can be of dissipative type (effective viscosities) or non-dissipative (e.g. effective sound velocity). 
The perturbative expression for this type of effective fluid parameters are governed by the non-linear mode couplings within the fluid and involve an averaging over the spectrum of perturbations \cite{Blas:2015tla}. In the field of heavy ion physics, similar concepts have been used before, for example the anomalous plasma viscosity of ref. \cite{Asakawa:2006tc} or the eddy viscosity of ref.\ \cite{Romatschke:2007eb}. 

At present, perturbative expressions of ref.\ \cite{Blas:2015tla} do not allow a clear distinction between effective viscosity and effective sound velocity. Only a linear combination of them is constrained by the growing scalar mode. Taking the resulting coarse-grained theory at some intermediate scale $k_m$ as an input, one can calculate the dark matter power spectrum in perturbation theory. A two-loop calculation agrees very well with numerical N-body simulations up to relatively small wavelengths and the convergence properties (tested by varying the matching scale) are somewhat better than for standard perturbation theory \cite{Blas:2015tla}. One can expect that viscous fluid dynamics will provide a useful approximation scheme for a statistical description of large scale structure in the coming years.

\subsection{Viscosity as a way to characterize dark matter}

In current cosmology codes, dark matter is often modeled as an ideal and pressure-less fluid. On the other side, we believe that dark matter consists of massive particles with small or vanishing interactions besides gravity. 
The question arises: what are the dissipative properties, in particular viscosities? Is there a way to constrain the dissipative properties from observation in order to learn more about dark matter, in particular about the so far rather unconstrained dark matter self-interaction?

In a sense, late-time cosmology faces in this respect a problem that is inverse to the main problem of heavy ion physics. The microscopic properties of QCD are already quite well known and with the experimental and theoretical effort of the heavy ion program, we are working on understanding the collective fluid properties of QCD better and better. Schematically,
$$\mathscr{L}_\text{QCD} \quad\to\quad \text{fluid properties}.$$
In late time cosmology, on the other hand, one can observe the fluid properties of dark matter via its coupling to gravity, for example the dark matter equation of state. By understanding them better, in particular also the dissipative properties, one can make some statements about the fundamental physics of dark matter. In the absence of a direct detection, the coupling to gravity via the energy-momentum tensor is actually the only way to access the Lagrangian of dark matter, although somewhat indirect,
$$\text{fluid properties} \quad\to\quad \mathscr{L}_\text{dark matter}.$$
It is needless so say that understanding the properties of dark matter is an important task and that insights from heavy ion physics on the detailed connection between microscopic physics and fluid properties (for the example of QCD) could be quite helpful.

In order to judge the current state of understanding of dark matter properties it is instructive to recall the limits on its self interaction. They come mainly from analyzing colliding galaxy clusters, most prominent the ``bullet cluster'' 1E0657-56. From the fact that no offset between the visible stars and the dark matter distribution - as inferred from gravitational lensing - is seen, one can derive upper bounds on the ratio of a potential dark matter self interaction cross section $\sigma$ and the dark matter particle mass $m_\text{DM}$. Typically, the constraints indicate that $\sigma/m_\text{DM}$ must be less than about one cm$^2/$g. Recently, an analysis of the galaxy cluster Abell 3827 has actually found a small offset between the dark matter distribution and the visible stars of a galaxy falling into the cluster \cite{Massey:2015dkw}. When interpreted as being due to a drag force mediated by dark matter self interactions that would indicate $\sigma/m_\text{DM}\approx 3$ cm$/$g \cite{Kahlhoefer:2015vua}. This interpretation for the observed offset is still under debate, however. In any case, these considerations show how little is known so far about dark matter self interactions and that one could potentially put interesting bounds by investigating the large scale fluid properties in more detail.

An interesting perspective in this regard is that precision cosmology could measure some information about the shear stress of dark matter direclty. Scalar excitations of gravity around an expanding Friedmann-Rorbertson-Walker background can be expressed in terms of two Newtonian potentials $\psi$ and $\phi$ (in conformal Newtonian gauge). Einsteins equations imply at linear order that their difference is related to the scalar part of the shear stress, i.\ e.\ the  part of the energy-momentum tensor that is orthogonal to the fluid velocity, symmetric and traceless,
$$
\left( \partial_i\partial_j - \tfrac{1}{3} \delta_{ij} \partial_k^2\right) (\phi-\psi) = 8\pi G_\text{N} a^2\; \pi_{ij}{\big |}_\text{scalar}.
$$
The scalar part of the shear stress can be expressed as the following combination of derivatives acting on some scalar field $\hat\pi$,
$$
\pi_{ij}{\big |}_\text{scalar} = \left( \partial_i\partial_j - \tfrac{1}{3} \delta_{ij} \partial_k^2\right) \hat \pi .
$$
Now, the Newtonian potential $\psi$ can be measured via its effect on the acceleration of matter. The combination $\psi+\phi$ can be measured by propagating light, for example via gravitational lensing or the Sachse-Wolfe effect. In this way one can -- a least in principle -- also access $\hat \pi$. Finally, the fluid velocity can be accessed via redshift space distortions and one can test the relation of $\hat \pi$ to fluid velocity gradients that is mediated by the shear viscosity and the corresponding relaxation time. One can expect that the foreseen missions in the next decade will provide many interesting and quantitatively precise results about the fluid properties of the cosmological large scale structure, for example the Euclid satellite which the European Space Agency plans to launch in 2020, see \cite{Amendola:2012ys}.

\subsection{Bulk viscosity}
If the bulk viscosity is non-vanishing, the bulk viscous pressure is negative for an expanding situation such as a fireball or the universe. For an isotropic and homogeneous universe one has in first order viscous fluid dynamics
$$
\pi_\text{bulk} = - \zeta \nabla_\mu u^\mu = - \zeta 3 H,
$$
where $H$ is the Hubble parameter.
It has been observed early that, if also the effective pressure, the sum of the thermodynamic pressure and bulk viscous pressure, would be negative, $p_\text{eff}=p+\pi_\text{bulk}<0$, this would act very similar to dark energy in Friedmann's equations \cite{Murphy:1973zz, Padmanabhan:1987dg, Fabris:2005ts, Li:2009mf, Velten:2011bg, Gagnon:2011id}. It could therefore lead to an accelerated cosmological expansion. It is, however, questionable whether negative effective pressure is a physically possible situation. In kinetic theory, the effective pressure is positive semi-definite. In the context of heavy ion physics, it was argued that an instability called ``cavitation'' could arise for $p_\text{eff}<0$ \cite{Torrieri:2007fb, Rajagopal:2009yw, Buchel:2013wxa, Habich:2014tpa, Denicol:2015bpa}. It is, however, not entirely clear what precisely happens at the instability and whether negative effective pressure could be metastable. 

Incidentally, cavitation could also play an interesting role for the fluid dynamics of heavy ion collisions. Bulk viscosity is expected to become sizable in the vicinity of the chiral crossover, see \cite{Denicol:2015bpa} for a recent discussion. A very interesting question is wether the associated negative bulk viscous pressure plays a role for the chemical and kinetic freeze-out taking place there. These considerations make clear that a more detailed understanding of bulk viscous effects and the instabilities associated with potentially negative effective pressure is urgently needed, both for heavy ion physics and for cosmology.

\subsection{The backreaction}
In a theoretical setup where one distinguishes between a background or expectation value and perturbations, one can also ask wether the dynamics of the background is affected by the perturbations. For example, in heavy ion physics, one can see an azimuthally symmetric initial energy density as background and the eccentricities $\varepsilon_m = \epsilon_m e^{im \Phi_m}$ as characterizations of perturbations around it. Within some approximation, harmonic flow coefficients $V_m=v_m e^{i m \psi_m}$ can be written in terms of linear and non-linear response to the eccentricities, e.\ g.\
$$
V_4 = \kappa_4^\text{L} \; \varepsilon_4 + \kappa_{4}^\text{NL}\; \varepsilon_2^2 + \ldots.
$$
However, the symmetries of the problem allow also non-linear contributions of perturbations to the background, which is here the azimuthally integrated particle spectrum,
$$
\frac{dN}{d\eta} = c + \kappa_{0,2}^\text{NL} \;\, \varepsilon_2^* \varepsilon_2 + \kappa_{0,3}^\text{NL} \;\, \varepsilon_3^* \varepsilon_3 + \kappa_{0,4}^\text{NL} \;\, \varepsilon_4^* \varepsilon_4 + \ldots
$$
There is in general no reason why the non-linear terms $\kappa_{0,2}^\text{NL}$ etc. should vanish. Their effect would be called a backreaction.

In cosmology, backreaction effects are particularly interesting because inhomogeneities could thereby affect the overall cosmological expansion. Einstein's equations are non-linear and it has been investigated whether this could have an influence on the the evolution of the universe, see \cite{Buchert:2011sx, Green:2013ica} for recent reviews. The question is not easy to answer, mainly because of technical difficulties with gravity as a gauge theory. Although there is no general consensus, most people believe now that gravitational backreaction is a rather small effect. In a recent work comparing fluid evolution equations describing heavy ion collisions to those describing the evolution of the universe \cite{Brouzakis:2014gka, Floerchinger:2014jsa}, we became aware of an additional backreaction effect that is due to dissipative terms, more specific shear and bulk viscosity. 

With this term taken into account, the cosmological evolution equation for the expectation value of the fluid energy density $\bar \epsilon = \langle \epsilon \rangle$ reads \cite{Floerchinger:2014jsa}
\begin{equation}
\tfrac{1}{a}\dot{\bar \epsilon} + 3 H  \left(\bar \epsilon + \bar p - 3 \bar \zeta H \right) =D
\label{eq:EnergyEqnBG}
\end{equation}
where the backreaction term is
\begin{equation}
\begin{split}
& D=  \tfrac{1}{a^2} \left\langle \eta \left[ \partial_i v_j \partial_i v_j + \partial_i v_j \partial_j v_i - \tfrac{2}{3} \partial_i v_i \partial_j v_j\right] \right\rangle\\
& \quad\quad + \tfrac{1}{a^2} \left\langle \zeta[\vec \nabla \cdot \vec v]^2 \right\rangle +
\tfrac{1}{a}\left\langle  \vec v \cdot \vec \nabla \left( p - 6 \zeta H \right) \right\rangle
\end{split}
\nonumber
\end{equation}
Note that $D$ vanishes for an unperturbed, homogeneous and isotropic universe but has contribution from shear viscosity (first term), bulk viscosity (second term) and thermodynamic work done by contraction against pressure gradients (third term) when inhomogeneities are present. The viscous terms contributing to $D$ are actually positive semi-definite because they are expectation values of squares multiplied by the shear and bulk viscosity, respectively. The above expression for $D$ was obtained for first order viscous fluid dynamics but can be generalized to higher orders or more general non-equilibrium situations. Note that for fluid velocity gradients comparable to the Hubble rate, $\tfrac{1}{a}\vec \nabla \vec v \sim H$, the backreaction term is of the same order as the bulk viscous pressure term in the background on the left hand side of Eq.\ \eqref{eq:EnergyEqnBG}. However, in contrast to the terms in the background, both shear and bulk viscosity contribute to $D$. Moreover, there can now be nontrivial effects that do not need negative effective pressure $p_\text{eff}$. Physically, the backreaction term accounts simply for the dissipated energy from the fluids macroscopic kinetic energy to internal energy.

An interesting question is how large $D$ would have to be in order to have a sizable effect onto the cosmological expansion. For this one needs an equation that determines the dynamics of the scale factor $a$. It is advantageous to use the trace of Einstein's equations, $R=8\pi G_\text{N} T^\mu_{\;\;\mu}$, because its spatial average does not depend on unknown quantities like e.\ g.\ $\langle (\epsilon + p_\text{eff}) u^\mu u^\nu \rangle$. One obtains
\begin{equation}
\tfrac{1}{a} \dot H + 2 H^2 = \tfrac{4\pi G_\text{N}}{3} \left(\bar \epsilon - 3 \bar p_\text{eff}\right).
\label{eq:traceEinsteinEqn}
\end{equation}
One can now analyze eqs.\ \eqref{eq:EnergyEqnBG} and \eqref{eq:traceEinsteinEqn} for given equation of state and backreaction term $D$. One finds for $p_\text{eff}=0$ that the backreaction is large enough to affect the expansion of the universe if the dimensionless ratio $4\pi G_\text{N} D/3H^3$ becomes of order one \cite{Floerchinger:2014jsa}. 

Based on estimates of the typical fluid velocity gradients one can now determine how large the viscosities would have to be in order to reach such a large backreaction term \cite{Floerchinger:2014jsa}. The corresponding values are rather large and it is currently not clear whether consistent models can be constructed where this is realized. In particular one would have to solve the evolution equations for the perturbations and background consistently to see whether large viscous effects can be brought into agreement with current knowledge about structure formation.

\section{Conclusions}
\label{}

In summary, there is an interesting and useful analogy between the physics of heavy ion collisions and cosmology. Both systems can be described by variants of relativistic fluid dynamics. For heavy ion collisions, the properties of this fluid are governed by QCD. At least in principle, enough information about the microscopic physics is known to determine the fluid properties theoretically. In contrast, the cosmological fluid at late times is governed by the microscopic physics of dark matter which one would like to understand better.

In this respect, insights into the connection between microscopic physics and fluid properties as they can be gained from the heavy ion program, are very useful for cosmology. Similarly, approximation schemes and theoretical concepts can be exchanged between the fields.

There are also common questions of relevance for both fields, for example whether negative effective pressure $p_\text{eff}=p+\pi_\text{bulk}<0$ is possible, whether there is an instability setting in at this point, and what consequences precisely this has.

Heavy ion physics can also learn from cosmology. In particular, much of the progress of cosmology in the last decades stems from a very detailed understanding of inhomogeneities or perturbations. The central theoretical tool that made this progress possible is cosmological perturbation theory. A very similar theoretical approach could be used for perturbations in the fluid dynamic description of heavy ion collisions and could complement the numerical simulations currently used. A first interesting benefit from this could be that initial conditions at the beginning of the fluid dynamic regime could be specified in a way that allows to concentrate on the most important wavelengths and to use a statistical description.

\section*{Acknowledgment}
I would like to thank Urs A. Wiedemann for many discussions and collaborations on the topics discussed here.





\bibliographystyle{elsarticle-num}



\end{document}